\documentclass[10pt,letterpaper]{article}

\usepackage{cogsci}

\cogscifinalcopy 

\usepackage{newtxtext,newtxmath}
\usepackage{apacite}
\usepackage{float} 
\usepackage{graphicx}
\usepackage{amsmath}
\usepackage{amsfonts}
\usepackage{booktabs}
\usepackage{makecell}

\title{Training-Driven Representational Geometry Modularization Predicts Brain Alignment in Language Models}

\author{
  {\large \bf Yixuan Liu (liuyixua22@mails.tsinghua.edu.cn)$^{\bf 1,2,*}$}\textbf{,}
  {\large \bf Zhiyuan Ma$^{\bf 1,2,*}$}\textbf{,}
  {\large \bf Likai Tang$^{\bf 1,2}$}\textbf{,} 
  {\large \bf Runmin Gan$^{\bf 1,2}$}\textbf{,}\\
  {\large \bf Xinche Zhang$^{\bf 1,2}$}\textbf{,}
  {\large \bf Jinhao Li$^{\bf 2,3}$}\textbf{,}
  {\large \bf Chao Xie (xiec199@gmail.com)$^{\bf 4,\dag}$}\textbf{,}
  {\large \bf Sen Song (songsen@tsinghua.edu.cn)$^{\bf 1,2,\dag}$} \\
  {$^{\bf 1}$School of Biomedical Engineering, Tsinghua University, Beijing, China} \\
  {$^{\bf 2}$Tsinghua Laboratory of Brain and Intelligence, Tsinghua University, Beijing, China} \\
  {$^{\bf 3}$School of Basic Medical Sciences, Tsinghua University, Beijing, China} \\
  {$^{\bf 4}$Department of Psychological and Cognitive Sciences, Tsinghua University, Beijing, China} \\
  {$^*$Equal Contributors }
  {$^\dag$Corresponding Authors}
}

\begin{document}

\maketitle

\begin{abstract}

How large language models (LLMs) align with the neural representation and computation of human language is a central question in cognitive science. Using representational geometry as a mechanistic lens, we addressed this by tracking entropy, curvature, and fMRI encoding scores throughout Pythia (70M–1B) training. We identified a geometric modularization where layers self-organize into stable low- and high-complexity clusters. The low-complexity module, characterized by reduced entropy and curvature, consistently better predicted human language network
activity. This alignment followed heterogeneous spatial-temporal trajectories: rapid and stable in temporal regions (AntTemp, PostTemp), but delayed and dynamic in frontal areas (IFG, IFGorb). Crucially, reduced curvature remained a robust predictor of model–brain alignment even after controlling for training progress, an effect that strengthened with model scale. These results links training-driven geometric reorganization to temporal-frontal functional specialization, suggesting that representational smoothing facilitates neural-like linguistic processing.

\end{abstract}

\section{Introduction}

Understanding how the human brain represents and computes language has long been a central challenge in cognitive neuroscience. The recent success of Transformer-based large language models (LLMs) has motivated their use as computational models of human language processing \cite{Biderman2023a,Chang2024a,Vaswani2017a}. Specifically, LLM activations could significantly predict fMRI responses to linguistic stimuli using linear encoding models \cite{AlKhamissi2025a,Caucheteux2022a,Goldstein2022a,Rathi2025a,Schrimpf2021a,Tuckute2024a}, suggesting that models and brains may share common principles of language representation and computation. 

Despite these successes, the mechanisms underlying model–brain alignment remain debated. While earlier work emphasized the role of syntactic structure and hierarchical processing \cite{AlKhamissi2025a,Caucheteux2023a,Murphy2014a}, other research suggested that alignment is primarily driven by event semantics and contextual meaning \cite{Kauf2023a}. Recent findings proposed a dissociation: models may rely on syntax for next-token prediction, yet their alignment with the brain appeared mediated by high-level semantic features \cite{Proietti2025a}. Beyond these linguistic distinctions, shared computational principles—specifically predictive coding—have been proposed as a unifying mechanism \cite{Goldstein2022a}. However, it remains unclear whether these effects reflect deep functional homology or merely a shared sensitivity to low-level statistical properties \cite{Hadidi2025a}, highlighting the limitations of explaining alignment solely through external linguistic features.

Given that high correlations may stem from shared sensitivity to low-level statistics rather than functional homology, focusing solely on external linguistic features may be insufficient. Fundamentally, encoding models map between a model’s representational space and the brain’s neural activity space, thus a mechanistic route is to probe the intrinsic geometry of representations. Recent work revealed that geometric properties—such as entropy and curvature—governed model adaptability \cite{Skean2025a}, consistent with peak neural alignment in intermediate layers that show distinct geometric signatures \cite{Caucheteux2022a}. Therefore, representational geometry was hypothesized to serve as a shared latent variable: if models and brains share computational principles, alignment should appear as structural consistency in their high-dimensional manifolds, not merely feature-level correlations. 

Crucially, these geometric structures emerge dynamically during training. Yet, most research relied on static, fully trained snapshots, overlooking the developmental trajectory of alignment. While recent studies have begun to explore training dynamics \cite{AlKhamissi2025a,Hosseini2024a}, they remain limited to coarse, network-wide averages. Consequently, fine-grained analyses at the level of specific regions (ROIs) and systematic investigations into layer-wise modularity remain absent.

\section{The Present Study}

Our study examines how representational geometry evolves with training and relates to brain alignment. We ask: \textbf{Does the training-induced differentiation of layers into geometric modules explain the fMRI predictability of the language network?} Specifically, we address three key sub-questions:
(1) \textbf{Emergence}: Does geometric modularity emerge and stabilize over training? (2) \textbf{Mapping}: How does fMRI predictability differ across modules and ROIs? (3) \textbf{Mechanism and Scale}: Within modules, do geometric metrics predict fMRI scores, and do these effects scale with model size?

To address these questions, we analyze the training dynamics of the Pythia model suite across multiple scales. We track the co-evolution of layer-wise representational geometry (entropy and curvature) and their fMRI encoding performance. Furthermore, we examine how these associations vary across model capacities (70M–1B) to determine if scale acts as a stabilizer for the geometry–alignment relationship.

Our main contributions are fourfold: (1) We show that training induces a stable geometric modularization of layers into low- and high-complexity modules. (2) We find a global alignment advantage for the low-complexity module, but with distinct ROI dynamics—early and stable in temporal regions, delayed and more dynamic in frontal regions. (3) We identify curvature as a robust predictor of fMRI encoding scores beyond training progress, with flatter representations yielding higher predictability. (4) We show that larger model scale stabilizes and strengthens geometry–alignment coupling, improving statistical detectability.

\section{Methods}


\subsection{Brain Data and Model}

\paragraph{Brain Data and Stimuli.}
We utilized the TUCKUTE2024 fMRI dataset \cite{Tuckute2024a}, which contains neural responses from 5 participants during the passive reading of 1,000 sentences. We analyzed five Regions of Interest (ROIs) in the left-hemisphere language network (netw): IFGorb, IFG, MFG, AntTemp, and PostTemp. For each sentence, responses were averaged across participants to maximize the signal-to-noise ratio.

\paragraph{Model.}
We employed the Transformer-based Pythia suite \cite{Biderman2023a} at four scales (70M, 160M, 410M, 1B). We analyzed 19 log-spaced checkpoints ranging from step 1 to 143K: 10 standard checkpoints for steps 1–512, followed by exponentially spaced checkpoints from step 1K onward (1K–128K), and the final checkpoint at step 143K.

\subsection{Representation Geometry Analysis}
Following prior work \cite{Skean2025a}, we characterized the layer-wise representational geometry by computing entropy and curvature. These metrics were calculated per sentence and averaged across stimuli.

\begin{figure}[h]
    \centering
    \includegraphics[width=\linewidth]{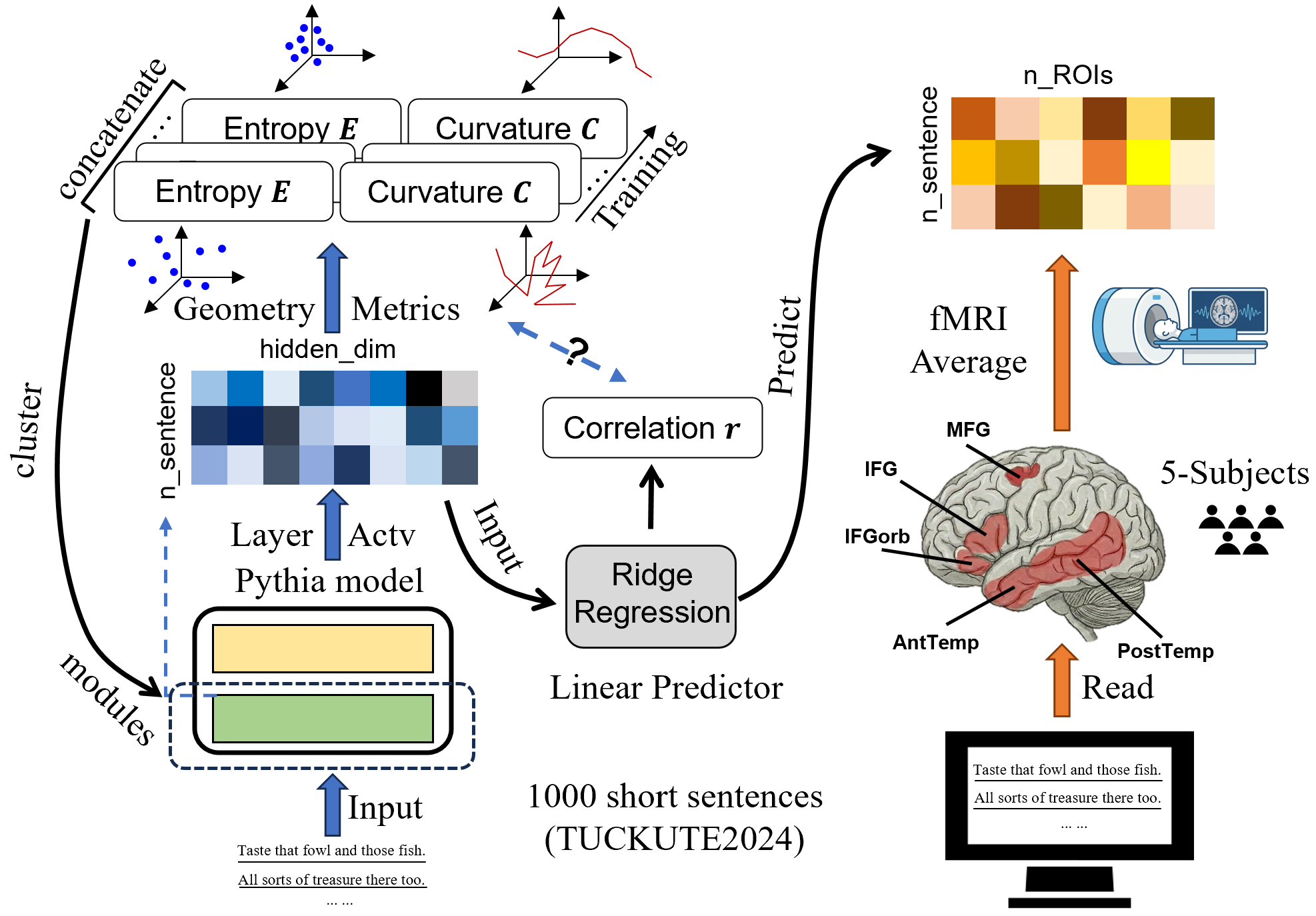}
    \caption{\textbf{Overview of the experimental framework.} We employ the TUCKUTE2024 dataset (1,000 sentences) to obtain both Pythia layer activations and human fMRI responses. Layer activations are mapped to subject-averaged fMRI data using ridge regression (5-fold CV). In parallel, we track representational geometry (entropy $E$ and curvature $C$) across training checkpoints to cluster model layers into modules, analyzing the relationship between these geometric module trajectories and fMRI predictability (Pearson correlation $r$).}
    \label{fig:method}
\end{figure}

\paragraph{Entropy.}
For sentence $s$ and layer $l$, let $\mathbf{Z}^{(l,s)}\in\mathbb{R}^{L_s\times d}$ be the matrix of token hidden states and $\mathbf{K}^{(l,s)}=\mathbf{Z}^{(l,s)}{\mathbf{Z}^{(l,s)}}^\top$ be the Gram matrix. Let $p_i=\lambda_i(\mathbf{K})/\mathrm{tr}(\mathbf{K})$ be defined over nonzero eigenvalues. We defined the von Neumann entropy as:
\begin{equation}
E^{(l,s)}=-\mathrm{tr}(\boldsymbol{\rho}\log\boldsymbol{\rho})=-\sum_i p_i\log p_i,\quad \boldsymbol{\rho}=\mathbf{K}/\mathrm{tr}(\mathbf{K}).
\end{equation}
Intuitively, $E^{(l,s)}$ quantifies spectral dispersion: low entropy indicates a compressed, low-rank representation dominated by few components, whereas high entropy reflects a more distributed, high-dimensional state.

\begin{figure*}
    \centering
    \includegraphics[width=\textwidth]{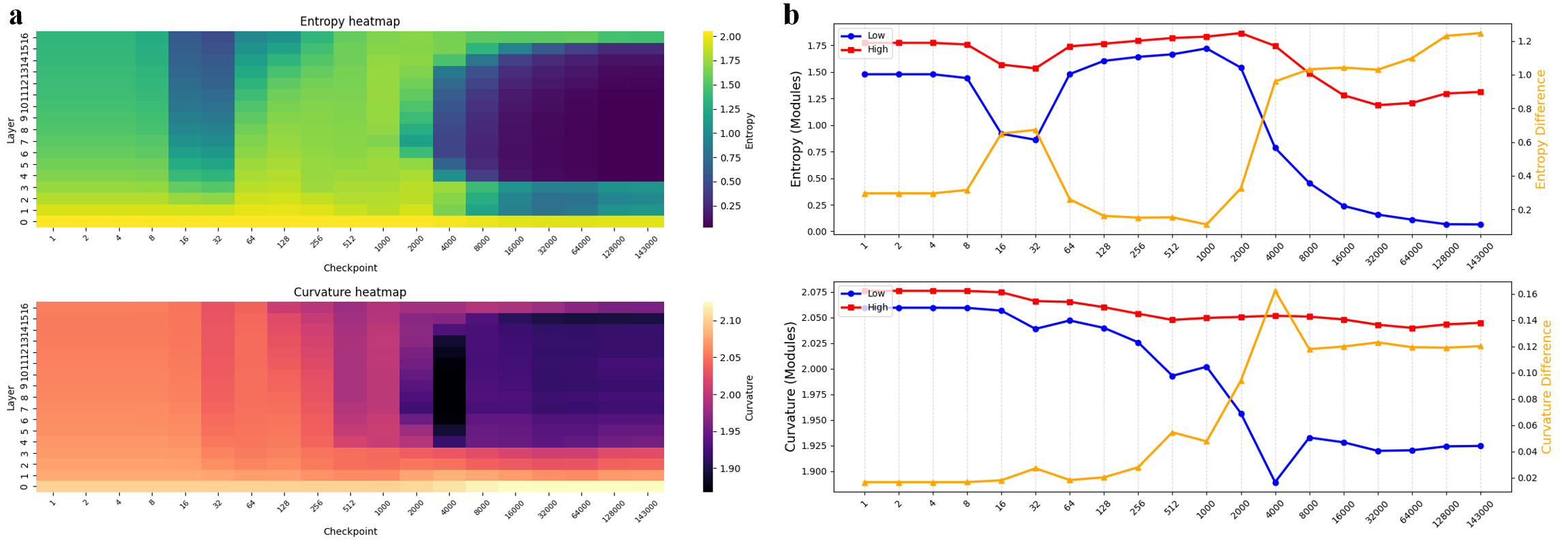}
    \caption{\textbf{Geometric evolution and modularization of Pythia-1B layers. (a)} Heatmaps of entropy (top) and curvature (bottom). The horizontal axis represents training checkpoints (log-scale, steps 1–143k), and the vertical axis corresponds to layers (0: embeddings, 1–16: Transformer blocks). Darker colors indicate lower values. \textbf{(b)} Aggregated geometry trajectories of the two identified layer clusters. The low-complexity module (blue) develops a stable low-entropy/curvature profile, while the high-complexity module (red) maintains higher values. The orange line shows the difference between the two modules.}
    \label{fig:geometry_emergence}
\end{figure*}

\paragraph{Curvature.}
Following \citeA{Hosseini2023a}, let $\mathbf{z}_k^{(l,s)}$ be the representation at token $k$ and $\mathbf{v}_k^{(l,s)}=\mathbf{z}_{k+1}^{(l,s)}-\mathbf{z}_k^{(l,s)}$. Curvature is defined as the mean turning angle:
\begin{equation}
C^{(l,s)}=\frac{1}{L_s-2}\sum_{k=1}^{L_s-2}\arccos\left(\frac{{\mathbf{v}_{k+1}}^\top \mathbf{v}_k}{\|\mathbf{v}_{k+1}\|\|\mathbf{v}_k\|}\right).
\end{equation}
Conceptually, $C^{(l,s)}$ measures the sharpness of token trajectories: higher curvature implies more local, fine-grained variation across tokens, whereas lower curvature indicates smoother, more globally consistent trajectories.

\paragraph{Layer-level Aggregation.}
We computed $E^{(l,s)}$ and $C^{(l,s)}$ for each stimulus sentence and averaged them across sentences to obtain layer-wise metrics $E_{t}^l$ and $C_{t}^l$ at each checkpoint $t$. This aggregation reduces content-specific variation and serves as a stable descriptor of geometry for clustering.

\subsection{Clustering and Robustness Analysis}
\paragraph{Geometry-Trajectory Features for Clustering.}
For each layer $l$, we constructed a geometry-trajectory vector by concatenating entropy and curvature across $T$ checkpoints. This vector captures how the layer’s geometry evolves over training:
\begin{equation}
\mathbf{x}_l = [E_{1}^l,C_{1}^l, \ldots,E_{T}^l,C_{T}^l].
\end{equation}

\paragraph{K-means Clustering and Robustness of Layer Modules.}
Motivated by the hypothesis of an interpretable low- vs. high-complexity modularization, we applied k-means ($K=2$, Euclidean) to $\{\mathbf{x}_l\}$ to identify two geometric modules. To verify that the partition was not driven by specific training stages, we assessed assignment stability in two ways: (i) checkpoint bootstrap, resampling checkpoints with replacement and reclustering; and (ii) leave-one-checkpoint-out (LOCO), dropping one checkpoint at a time and reclustering. Stability was quantified per layer as the fraction of runs in which the layer retained its original cluster label.

\begin{figure*}
    \centering
    \includegraphics[width=\textwidth]{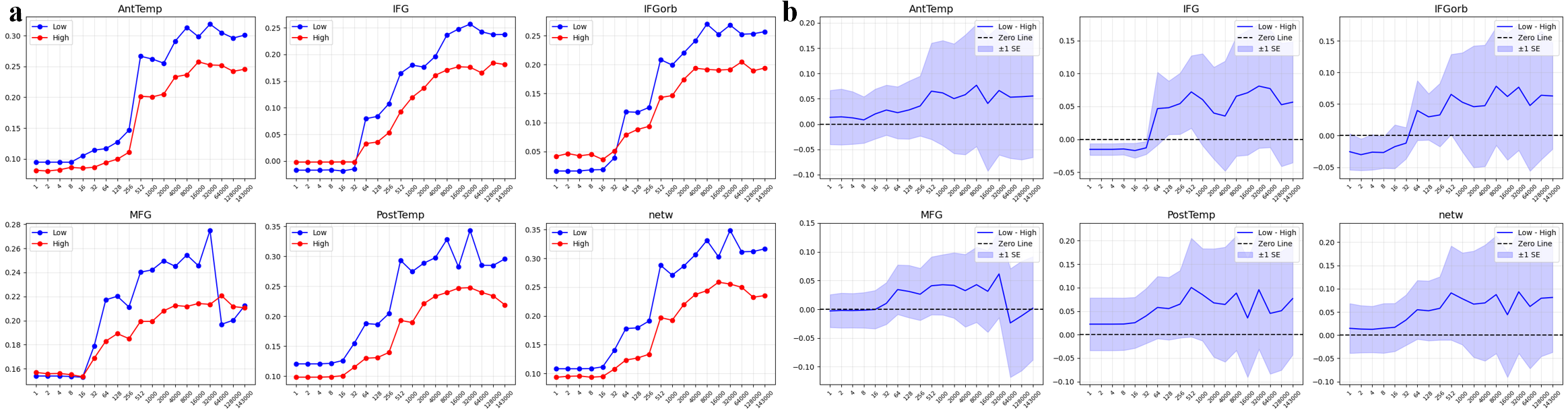}
    \caption{\textbf{fMRI alignment time course by geometry module.} \textbf{(a)} Mean cross-validated fMRI encoding scores (Pearson correlation) for the low-complexity module (blue) and high-complexity module (red) across six language ROIs over training checkpoints. \textbf{(b)} Module gap $\Delta F$ over training (blue line: mean across layers; shaded: SD across layers at each checkpoint). The dotted line denotes zero. Temporal ROIs show an earlier, more stable advantage for the low-complexity module, whereas frontal ROIs exhibit a more gradual and dynamic separation.}
    \label{fig:fmri-timecourse}
\end{figure*}

\paragraph{Geometry--fMRI Coupling Analysis.}
We examined both trajectory-level and conditional associations. Trajectory coupling was quantified by Pearson correlations across checkpoints between module-averaged geometry and encoding scores. Conditional coupling was assessed using ROI-specific fixed-effects regressions:
$$fMRI^{(r)}_{l,t} = \beta_G\,G_{l,t} + \beta_t \log t + \alpha_l + \varepsilon_{l,t},$$
where $G_{l,t}\in\{E_{l,t},C_{l,t}\}$ and $\alpha_l$ denote layer fixed effects. To improve comparability across model scales, we z-scored $fMRI^{(r)}_{l,t}$ within each ROI and z-scored $G_{l,t}$ within each module before estimating $\beta_G$. We used checkpoint-clustered robust standard errors (SEs) and applied Benjamini-Hochberg-FDR (BH-FDR) correction across ROIs $(q<0.05)$.

\section{Results}
\subsection{Emergence of Geometry-Based Layer Modules Across Training}
We first characterize the geometric evolution of Pythia-1B representations throughout the training dynamics. As shown in \textbf{Fig.~\ref{fig:geometry_emergence}a}, geometry is relatively homogeneous early on, but a clear banded differentiation emerges in mid-to-late training (steps \(>\)4k): middle-to-upper layers converge to a stable low-entropy pattern with a concurrent reduction in curvature, whereas shallow layers remain comparatively high on both metrics. This increasing inter-layer heterogeneity is significant: the inter-layer IQR rises from 0.31 to 0.81 for entropy and from 0.015 to 0.060 for curvature (permutation over checkpoints, \(p=0.001\) for both).

Motivated by this structure, we clustered per-layer geometric trajectories (K-means, \(K=2\)) and obtained two robust layer modules. Assignments were stable under Leave-One-Checkpoint-Out (LOCO) evaluation (all layers = 1.0) and bootstrap resampling (minimum stability \(\ge 0.92\)). Aggregating layers within each cluster (\textbf{Fig.~\ref{fig:geometry_emergence}b}) shows that \textbf{low-complexity module (layer 4-15)} consistently exhibits lower entropy and curvature than \textbf{high-complexity module (other layers)} (mean Cohen's \(d=2.42\) and \(d=2.64\), respectively; FDR-corrected \(p<0.05\) across checkpoints). Entropy follows a two-phase pattern (a transient dip at steps 8--64 and a larger decline after \(\ge\)1k), whereas curvature decreases more continuously; correspondingly, the inter-cluster gap expands and stabilizes late in training (step \(>\)4k).

Overall, Pythia-1B develops a repeatable geometry-based modularization, providing a principled layer grouping for downstream analyses of module-specific brain alignment.

\subsection{Differential Brain Alignment of Geometry Modules Across Language ROIs}
Having established geometry-based layer modularization, we asked whether the two modules differ in fMRI encoding performance across language ROIs, and whether the emergence of this difference is ROI-specific. For each checkpoint $t$, we defined the module-averaged fMRI encoding score within each ROI and the alignment gap as
$\Delta F(t) = \text{Score}_{\text{Low}}(t) - \text{Score}_{\text{High}}(t)$, and evaluated its magnitude and temporal dynamics.

\subsubsection{Global Trend: Low-Complexity Modules Exhibit a Robust Alignment Advantage}
Across all ROIs, encoding scores increase with training (\textbf{Fig.~\ref{fig:fmri-timecourse}a}). low-complexity module consistently outperforms high-complexity module, yielding significantly positive $\Delta F(t)$ across the full trajectory in every ROI (paired tests on $\Delta F(t)$; all $p<0.005$, FDR-corrected). The gap is largest in Temporal ROIs (PostTemp: $d=2.09$; AntTemp: $d=1.87$) and moderate in Frontal ROIs and MFG (IFG: $d=0.95$; IFGorb: $d=0.76$; MFG: $d=0.79$). Thus, lower-complexity representations show stronger biological correspondence across the language network.

\subsubsection{ROI Differentiation: Timing and Trajectory of Advantage Establishment}
Although low-complexity module is globally favored, the onset and shape of this advantage vary across ROIs (\textbf{Fig.~\ref{fig:fmri-timecourse}b}).

\begin{figure*}
    \centering
    \includegraphics[width=1\linewidth]{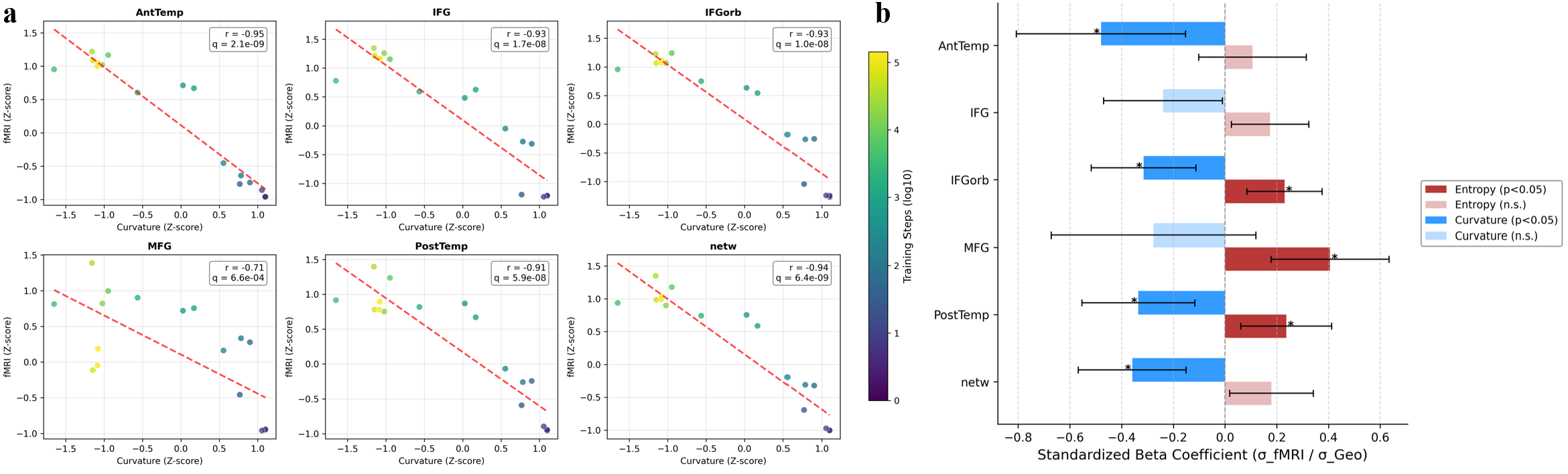}
    \caption{\textbf{Geometric co-evolution and conditional coupling with brain alignment.} \textbf{(a)} Scatter plots relating mean geometric metrics (Curvature/Entropy, averaged across layers) within low-complexity module to fMRI encoding scores across checkpoints ($n=19$). \textbf{(b)} Standardized regression coefficients ($\beta_G$) and 95\% CIs from $fMRI \sim G + \log(t) + \alpha_l$, where $G$ is z-scored Curvature or Entropy and $\alpha_l$ denotes layer fixed effects. Robust SEs are clustered by checkpoint; significance is based on BH--FDR correction across 6 ROIs for each metric ($q<0.05$).}
    \label{fig:geometry-fmri}
\end{figure*}

\paragraph{(i) Temporal ROIs (AntTemp / PostTemp): Early-onset and Stable.}
Both AntTemp and PostTemp exhibit significantly positive $\Delta F(t)$ from the earliest checkpoints (step $\le 64$; $t>5.8, p<0.001$), with no sign reversal. AntTemp additionally shows a sharp increase, with a change point at step 512 ($p<0.05$) aligned with the onset of a stable plateau, indicating rapid and persistent specialization.

\paragraph{(ii) Frontal ROIs (IFG / IFGorb): Gradual Differentiation with Early Inversion.}
Frontal ROIs separate earlier via detectable transition (change point at step 64, $p<0.01$), but the gap evolves more dynamically thereafter. Notably, IFGorb shows a brief early tendency toward high-complexity module (step $\le 64$: mean $\Delta F=-0.014$, $p\approx 0.09$), before switching to a sustained low-complexity module advantage.

\paragraph{(iii) MFG: Small and Unstable Module Specificity.}
MFG shows the weakest and least consistent gap (mean $\Delta F\approx 0.018$; $t=3.44$). No reliable early or mid-stage change point is detected (only a late signal near step 64k), and the nominal emergence point (step 32) is not followed by a stable plateau, suggesting limited selectivity for either module.

\paragraph{(iv) Network-level (lang\_LH\_netw): Aggregate Robustness.}
At the language-network level, the low-complexity module advantage is strong and consistent ($t=8.16, d=1.87$), stabilizing around step 512, consistent with module-level geometric differentiation translating into macroscopic encoding gains.

\subsection{Dynamic Analysis During Training: The Evolutionary Relationship Between Geometric Features and fMRI Encoding Scores}

In this section, we examine how representational geometry co-evolves with fMRI encoding performance during training, and whether geometric metrics retain explanatory power after controlling for training progress and stable layer differences.

\subsubsection{Co-evolution on the Training Trajectory}
We first quantify \textbf{trajectory-level coupling} between geometry and alignment by correlating checkpoint-averaged geometric metrics with checkpoint-averaged fMRI scores (one point per checkpoint, $n=19$). As shown in \textbf{Fig.~\ref{fig:geometry-fmri}a}, \textbf{Curvature} exhibits a strong negative correlation with fMRI scores across ROIs: with the exception of MFG ($r=-0.71, p<0.001$), all ROIs show $|r|>0.91$ ($p<0.001$). Thus, as training proceeds, curvature decreases while encoding performance increases, yielding tightly coupled trajectories. \textbf{Entropy} shows a weaker overall negative correlation with fMRI scores ($r\approx-0.60$, not shown), and is not significant in MFG ($p>0.30$). Together, these results establish a strong co-evolution between geometric “flattening” and improved brain alignment.

\subsubsection{Conditional Coupling Controlling for Training Progress}

Trajectory coupling can largely reflect shared dependence on training time. We therefore test \textbf{conditional coupling} within the low-complexity module ($N=228$ per ROI) using the ROI-specific regression model defined in Methods (Geometry--fMRI coupling analysis). As illustrated in \textbf{Fig.~\ref{fig:geometry-fmri}b}, after controlling for training progress, \textbf{Curvature} remains a robust negative predictor of absolute alignment: AntTemp, IFGorb, PostTemp, and the network average show significantly negative coefficients ($\beta\in[-0.050,-0.032]$, BH--FDR $q<0.05$), with a similar negative trend in IFG. In contrast, \textbf{Entropy} shows ROI-dependent conditional effects, with significant positive coefficients in IFGorb, MFG, and PostTemp ($\beta\in[0.021,0.023]$, $q<0.05$) but non-significant effects in AntTemp and IFG. Note that these estimates are conditional on the training trend, their signs need not match the raw trajectory correlations. Overall, curvature exhibits stable conditional coupling across ROIs, whereas entropy is more spatially heterogeneous.

\subsubsection{Scale Effect}

To examine the robustness of these geometry-brain relationships across scales, we extended the analysis to Pythia models from 70M to 1B, (\textbf{Table~\ref{tab:cross_scale_stacked}}).\par\leavevmode

\begin{table}[t]
    \centering
    \scriptsize 
    \setlength{\tabcolsep}{1.5pt} 
    
    \caption{\textbf{Cross-Scale Statistics.} Summary of Curvature ($C$) and Entropy ($E$) coupling with fMRI. $r$: median correlation (range). $\beta$: median regression coefficient (range). $N_{-/+}$: number of ROIs with $\beta < 0$ (for $C$) or $\beta > 0$ (for $E$). Det.: ROIs passing FDR correction ($q < 0.05$).}
    \label{tab:cross_scale_stacked}
    \vspace{3mm}
    
    \begin{tabular}{l cccc c cccc}
        \toprule
        & \multicolumn{4}{c}{\textbf{Curvature ($C$)}} & & \multicolumn{4}{c}{\textbf{Entropy ($E$)}} \\
        \cmidrule(lr){2-5} \cmidrule(lr){7-10}
        \textbf{Scale} & $r$ & $\beta$ & $N_{-}$ & Det. & & $r$ & $\beta$ & $N_{+}$ & Det. \\
        \midrule
        70M & \makecell{-0.785 \\ \tiny(-0.83, -0.71)} & \makecell{0.021 \\ \tiny(-0.06, 0.30)} & 2 & 1/6 & & \makecell{-0.576 \\ \tiny(-0.60, -0.51)} & \makecell{0.230 \\ \tiny(0.08, 0.36)} & 6 & 5/6 \\
        \addlinespace[3pt]
        160M & \makecell{-0.838 \\ \tiny(-0.90, 0.50)} & \makecell{0.127 \\ \tiny(-0.07, 0.33)} & 1 & 0/6 & & \makecell{-0.567 \\ \tiny(-0.68, 0.42)} & \makecell{0.184 \\ \tiny(0.03, 0.32)} & 6 & 3/6 \\
        \addlinespace[3pt]
        410M & \makecell{-0.912 \\ \tiny(-0.95, -0.78)} & \makecell{-0.150 \\ \tiny(-0.27, 0.00)} & 5 & 1/6 & & \makecell{-0.648 \\ \tiny(-0.77, -0.40)} & \makecell{0.124 \\ \tiny(-0.05, 0.34)} & 5 & 2/6 \\
        \addlinespace[3pt]
        1B & \makecell{-0.931 \\ \tiny(-0.95, -0.71)} & \makecell{-0.325 \\ \tiny(-0.48, -0.24)} & 6 & 4/6 & & \makecell{-0.603 \\ \tiny(-0.65, -0.25)} & \makecell{0.205 \\ \tiny(0.11, 0.41)} & 6 & 3/6 \\
        \bottomrule
    \end{tabular}
\end{table}
Trajectory coupling between curvature and fMRI scores is  negative across sizes (median $r$ from $-0.785$ in Pythia-70M to $-0.931$ in Pythia-1B, only $r>0$ in MFG of Pythia-160M), suggesting a scale-general synchrony of decreasing curvature and increasing alignment. Entropy shows a similar co-variation across scales, but with weaker and less consistent effect sizes than curvature.

In contrast, conditional coupling strengthens with scale. In smaller models (70M/160M), $\beta(\text{Curv})$ is small with weak sign consistency and limited detectability (0--1 FDR-significant ROIs). The effect turns negative at 410M (median $\beta=-0.15$) and becomes stronger and more consistent at 1B (median $\beta\approx-0.325$; 6/6 ROIs negative; 4/6 FDR-significant). Entropy shows positive conditional effects but does not exhibit a comparable scale-dependent increase in detectability. Thus, increasing capacity selectively stabilizes low curvature as a robust conditional predictor of improved brain alignment.

Overall, these findings reveal a distinct scale effect: increasing model capacity selectively stabilizes low curvature as a robust conditional predictor of improved brain alignment, whereas the entropy--alignment relationship remains comparatively heterogeneous and does not show the same scale-dependent strengthening.

\section{Discussion}

Our study establishes representational geometry as a critical intermediate phenotype linking training dynamics to model--brain alignment. We demonstrate that geometric modularization—the spontaneous differentiation of layers into stable low- and high-complexity clusters—is a functional reorganization driven by learning, not an architectural artifact. The consistent alignment advantage of the low-complexity module implies that the brain favors representations compressed into lower-dimensional manifolds, facilitating linear readout. Crucially, the distinct trajectories of curvature (monotonic smoothing) versus entropy (non-monotonic dynamics) suggest they index different learning phases: curvature reflects progressive manifold smoothing, while entropy captures the tension between early adaptation and late-stage structural compression.

\subsection{Spatiotemporal Heterogeneity in ROI Alignment}
A key finding is the dissociation between Temporal and Frontal alignment trajectories. Temporal ROIs (AntTemp, PostTemp) show an early-onset and non-inverting preference for the low-complexity module, whereas frontal ROIs (IFG, IFGorb) exhibit a delayed re-configuration and even a transient early inversion. This dissociation reflect distinct computational demands on representational stability versus controlled synthesis.

Temporal regions (AntTemp, PostTemp) primarily support mental-lexicon retrieval and basic semantic mapping, computations that benefit from stable, low-ambiguity codes across contexts. During training, the model provides such stability early by compressing representations into a low-complexity space, so temporal alignment “locks in” quickly and remains non-inverting.

In contrast, frontal regions (IFG, IFGorb) support unification and control—combining retrieved units into context-dependent structures and resolving competition—which require flexible interaction patterns that typically emerge and stabilize later in training. As a result, frontal alignment is delayed and best understood as a computational dependency: high-level synthesis cannot settle while the underlying semantic substrate is still shifting, so frontal ROIs effectively wait for the Temporal low-complexity foundation to stabilize, with transient early inversions plausibly reflecting this unsettled regime.

\subsection{Curvature as a Mechanistic Bridge and the Entropy Puzzle}
Our analysis isolates curvature as a robust, mechanistic explanatory variable for alignment.
Curvature exhibits a consistent negative coupling with fMRI scores both globally (across the training trajectory) and locally (conditional on training progress), suggesting that manifold smoothness is a universal prerequisite for linear mapping to neural responses.

In contrast, entropy presents a puzzle: while globally negatively correlated with alignment, it often shows a positive conditional coefficient within specific training stages.

This apparent contradiction points to a non-monotonic relationship.
Globally, the reduction of entropy reflects the necessary formation of structure; however, locally—once a baseline structure is established—slightly higher entropy may index a richer effective dimensionality or capacity that captures more variance in neural data.This implies that entropy is a "double-edged sword" balancing compression and capacity, whereas curvature represents a more unambiguous axis of representational quality where "flatter is better" for biological alignment.

\subsection{The Role of Model Scale}
The relationship between geometry and alignment exhibits a clear scale effect.
While the global trajectory coupling is evident across all model sizes, the conditional coupling—specifically the robust negative effect of curvature—becomes statistically detectable and consistent only in larger models (e.g., Pythia-1B).
Thus, scale stabilizes the model’s geometric regime. Larger models are more likely to converge to the "brain-like" low-curvature state with sufficient signal-to-noise ratio, thereby rendering the subtle, within-stage geometric effects detectable against the background of training noise.
Thus, scale acts as an amplifier that reveals the underlying mechanistic link between manifold structure and neural encoding.

\subsection{Limitations and Future Directions}
Our findings should be interpreted within specific boundary conditions. First, the reliance on linear encoding models may underestimate alignment in non-linear computational regions (e.g., Frontal cortex), and our correlational design does not yet establish causality. Second, our conclusions are derived from the Pythia suite under a passive reading task. Future work should address these gaps by: (1) employing non-linear readouts (e.g., kernel ridge regression) to verify readout-dependence; (2) testing generalization across diverse architectures and active tasks; and (3) performing causal interventions—such as regularizing for curvature during training—to definitively test whether geometric smoothing drives brain alignment.

\section{Conclusion}
We investigated how representational geometry evolves during training and relates to model–brain alignment. Across Pythia checkpoints, layers self-organize into low-/high-complexity modules, and the low-complexity module consistently yields higher fMRI encoding across language ROIs, with early stable temporal selectivity and more gradual frontal differentiation. Curvature remains negatively associated with encoding after controlling for training progress and layer effects, and the geometry–alignment link becomes more detectable (and conditionally stronger) with scale. Together, these results identify training-driven geometric modularization as a mechanistically informative axis for explaining and potentially predicting model–brain alignment, offering a complementary lens of geometry beyond traditional linguistic perspectives to decipher the underlying mechanisms. Future work should test causal interventions, go beyond linear readouts, and assess generalization across architectures and neural modalities.

\bibliographystyle{apacite}

\setlength{\bibleftmargin}{.125in}
\setlength{\bibindent}{-\bibleftmargin}

\bibliography{refs}
\nocite{Fresen2024a}
\nocite{oota2022long}
\nocite{urbina2026brain}
\nocite{yetman2024representation}
\nocite{hefar}
\nocite{tuckute2024language}
\nocite{pasquiou2023information}

\end{document}